\documentstyle[prl,aps]{revtex}
\tighten
\begin{document}
\draft
\twocolumn[
\hsize\textwidth\columnwidth\hsize\csname @twocolumnfalse\endcsname

\title{
DC and AC Josephson Effect in a Superconductor-Luttinger
Liquid-Superconductor System
}
\author{Rosario Fazio$^{(1,2)}$, F. W. J. Hekking$^{(3)}$,
	and A. A. Odintsov$^{(4,5)}$}
\address{
 $^{(1)}$Institut f\"ur Theoretische Festk\"orperphysik,
Universtit\"at Karlsruhe, 76128 Karlsruhe, Germany\\
 $^{(2)}$
	  Istituto di Fisica, Universit\`a di Catania, viale A. Doria 6,
	  95129 Catania, Italy $^{\ast}$\\
 $^{(3)}$Theoretical Physics Institute, University of Minnesota,
116 Church Street SE, Minneapolis, MN 55455\\
 $^{(4)}$Department of Applied Physics, Delft University of Technology,
P.O. Box 5046, 2600 GA Delft, The Netherlands\\
 $^{(5)}$
  Nuclear Physics Institute, Moscow
State University, Moscow 119899 GSP, Russia $^{\ast}$ \\
}
\date{draft, \today}
\maketitle

\begin{abstract}
We calculate both the DC and the AC Josephson current through a one-dimensional
system of interacting electrons, connected  to two superconductors by tunnel
junctions.  We treat the (repulsive) Coulomb interaction  in the framework of
the one-channel, spin-$1/2$ Luttinger model.\\
The Josephson current is obtained for  two geometries of experimental
relevance:
a quantum wire and a ring. At zero temperature, the critical current is found
to  decay algebraically  with increasing distance $d$ between the junctions.
The
decay is characterized by an exponent which  depends on the strength of the
interaction. At finite temperatures $T$, lower than the superconducting
transition temperature $T_c$, there is  a crossover from algebraic to
exponential decay of the critical current as a function of $d$,  at a
distance of the order of
$\hbar v_F/k_B T$. Moreover, the dependence of critical current on temperature
shows non-monotonic behavior.\\
If the Luttinger liquid is confined to a ring of circumference $L$,  coupled
capacitively to a gate voltage and threaded by a magnetic flux, the Josephson
current shows remarkable parity effects under the variation of these
parameters.
For some values of the  gate voltage and applied flux, the ring acts as a
$\pi$-junction. These features are robust against thermal fluctuations up to
temperatures on the order of $\hbar v_F/k_B L$.\\
For the wire-geometry, we have also studied the AC-Josephson effect.  The
amplitude and the phase of the time-dependent Josephson current are affected by
electron-electron interactions. Specifically, the amplitude shows pronounced
oscillations as a function of the bias voltage due to the difference between
the velocities of spin and charge excitations in the Luttinger liquid.
Therefore, the AC Josephson effect can be used as a tool for the observation of
{\em spin-charge} separation.
\end{abstract}

]

\narrowtext

\section{Introduction}
Due to the recent development of superconductor-semiconductor (S-Sc)
integration technology it has become possible to observe the transport of
Cooper
pairs through S-Sc mesoscopic interfaces. Examples are the supercurrent through
a two-dimensional electron gas (2DEG)  with Nb contacts (S-Sc-S
junction)~\cite{Nitta} and   excess low-voltage conductance due to Andreev
scattering in Nb-InGaAs (S-Sc)  junctions~\cite{Kastalsky}. The transfer of
{\em
single} electrons through the interface between a semiconductor and a
superconductor with energy gap $\Delta$ is exponentially suppressed at
low temperatures  and bias voltages $k_BT,eV \ll \Delta$ ($e$ is the electron
charge). Instead, electrons will be transferred {\em in pairs} through the
interface, a phenomenon known as Andreev-reflection~\cite{Andreev}. It has been
realized only recently, that the  phase coherence between the two electrons
involved in this process could give rise to distinct signatures in the
transport
properties of mesoscopic S-Sc-S and S-Sc systems~\cite{Beenakkerrev,NATO-ARW}.

If the normal (Sc) region is free of disorder, the propagation of electrons is
ballistic. Phase coherence between the two electrons is maintained over the
length $L_{\rm cor} = \hbar v_F/ \mbox{max } \{ k_BT,eV \}$, where
$v_F$ is the Fermi velocity. In this regime, the critical current
$I_c$ through a short and narrow constriction in a high-mobility
non-interacting 2DEG, connected to two superconductors should be
quantized~\cite{Beenakkerquant}: each propagating mode contributes an amount
$e\Delta /\hbar$ to the critical current.

In the presence of disorder in the normal region, the motion of the two
electrons will be diffusive. Like in disordered metals, the phase coherence
between the two electrons is limited by the correlation length
$L_{\rm cor} = \sqrt{ \hbar D/ \mbox{max } \{ k_BT,eV \}}$,  where $D$ is the
diffusion constant. For instance, the excess low-voltage conductance in S-Sc
junctions~\cite{Kastalsky} can be explained in terms of constructive
interference occurring over this length scale between the two electrons
incident on the S-Sc interface~\cite{vanWees,Hekking}.

In these examples, electron-electron interactions are neglected. It
is well known, though, that they may have a strong influence on the transport
properties of mesoscopic systems. In general the interactions modify the
phase-coherence length $L_{\rm cor}$,  which poses limitations on the
abovementioned mesoscopic effects. In specific cases the effects of
electron-electron interactions  will strongly depend on the layout of the
system under consideration.

For example, the interactions will modify the critical current $I_c$ through a
normal metallic slab sandwiched between  two superconductors~\cite{Aslamasov}.
If
the coupling between normal metal and superconductors is weak (tunneling
regime)
and the size of the slab (and hence its electric capacitance
$C$) is small,  a phenomenological capacitive model~\cite{Grabert} can be used
to describe the effect of interactions. As a  result the critical current shows
strong resonant dependence on  the electro-chemical potential of the slab
dependence has different character  for $E_C < \Delta$ and $E_C > \Delta$, $E_C
= e^2/2C$ being the  charging energy.  On the other hand, if the normal metal
and the superconductor are well coupled (regime of Andreev reflection),
electron-electron interactions will modify the results obtained in
Ref.~\onlinecite{Aslamasov} in quite a different fashion. A perturbative
treatment of the interactions~\cite{Khmelnitskii} shows that an additional
supercurrent through the slab arises, whose sign depends on the nature of the
interactions in the slab (attractive or repulsive), and whose phase-dependence
has period $\pi$ (rather than $2\pi$ in the non-interacting case).

If instead of a metal a low-dimensional Sc nanostructure with a small electron
concentration is  considered, the abovementioned descriptions of the
electron-electron interactions are no longer sufficient. In one-dimensional
(1D)
systems the Coulomb interactions can not be treated as a weak perturbation.  As
a result a non-perturbative, microscopic treatment of  interactions is
required.
For 1D systems this can be done in the framework of the  Luttinger
model~\cite{Emery}. Interactions have a  drastic consequence: there are no
fermionic quasiparticle excitations. Instead, the low energy excitations of the
system consist of independent long-wavelength oscillations of the charge and
spin density, which propagate with different velocities. The density of states
has power-law asymptotics at low energies and the  transport properties cannot
be described in terms of the  conventional Fermi-liquid approach. For a quantum
wire with an arbitrarily small barrier this leads to a complete supression of
transport at low energies~\cite{Kane,MatveevGlazman}.

Another interesting feature arises in 1D interacting systems of a finite size.
For a Luttinger liquid confined to a ring, Loss~\cite{Loss} found remarkable
parity effects~\cite{footnote1} for the persistent currents. He used the
concept
of Haldane's topological excitations~\cite{Haldane}, extending the previous
work
of Byers and Yang for non-interacting electrons in a ring~\cite{Byers}.
Depending on the parity of the total number of electrons on the ring, the
ground
state is either diamagnetic or paramagnetic. For spin-1/2 electrons an
additional sensitivity on the electron number {\em modulo} 4 has been
found~\cite{Loss2}. Experimental evidence for Luttinger liquid behavior
in Sc nanostructures has been found recently. The dispersion of separate spin
and charge excitations in GaAs/AlGaAs quantum wires has been measured with
resonant inelastic light scattering~\cite{Goni}. Transport measurements on
quantum wires have revealed power-law dependence of the conductance as a
function  of temperature~\cite{Tarucha}.

In view of this we expect that electron-electron interactions may well have
drastic, observable consequences in systems which consist of low-dimensional Sc
nanostructures connected to superconductors. In this paper, we will study the
Josephson current through a Luttinger liquid~\cite{Fazio,Maslov}. Specifically,
we consider two geometries which can be realized experimentally: a long wire
with
contacts to two superconductors at a distance $d$ (see Fig.~\ref{system}a) or a
ring shaped Luttinger liquid  shown in Fig.~\ref{system}b. In both cases the 1D
electron liquid is connected to the superconducting electrodes  by tunnel
junctions. This is an interesting system from various points of view. First of
all, it enables one to study in a microscopic way how the Coulomb interaction
influences the phase-coherent propagation of {\em two} electrons through a 1D
normal region~\cite{Fisher}. Secondly, various  aspects of transport in
mesoscopic systems (parity effects and interference combined with
electron-electron interactions) and their interplay can be  enlightened in such
a device. Finally, since the Josephson effect is a ground state property, the
Josephson current can be used as a tool to probe the ground state of an
interacting electron system. In particular, for the ring geometry in the
presence
of an Aharonov-Bohm flux, the various possible ground-state configurations can
be determined by studying flux and gate voltage dependence of the critical
current.

The paper is organized as follows. In Section 2 we briefly review the
properties
of the spin-1/2  Luttinger model. In Section 3 the general formalism  for the
DC-Josephson effect is presented. The DC-Josephson current is obtained by
evaluating the contribution to the free energy  which depends on the difference
of the superconducting phases. Starting from the general
expression~(\ref{freeenergy}) we then consider various interesting limiting
situations. In Section 4, the wire geometry  is considered (see Fig. 1a). The
critical current decays as a power of the distance $d$ between the contacts.
The
exponent depends on the interaction strength. We distinguish two cases in which
the characteristic energy $\hbar v_F /d$   for the 1D system is either  much
smaller or much larger than the superconducting gap $\Delta$. For the ring
geometry (Section 5, see Fig. 1b), we focus on the dependence of the  critical
current on the applied gate voltage and/or flux. Both Section 4 and 5 contain a
discussion of the effect of finite temperatures. Section 6 is devoted to the
AC-Josephson effect. In this case the imaginary time approach of the previous
sections is inadequate and we will use a real time formulation.  The amplitude
of the AC component is found to show oscillations as a function of voltage due
to spin-charge separation. In the last section we present the conclusions.\\

\section{The Spin-1/2 Luttinger Liquid}
We start the description of the model we use by reviewing the theory of 1D
interacting spin-1/2 fermions (throughout, we use $\hbar = k_B = 1$). The
long-wavelength behaviour of such a quantum wire of length $L$ is  governed by
the Hamiltonian~\cite{Haldane}
\begin{equation}
	\hat{H}_L
	=
	\int \limits _{-L/2} ^{L/2}\frac{dx}{\pi }
	\sum _{j}
	v_j
	\left[
		\frac{g_j}{2} (\nabla \phi _j )^2 +
		\frac{2}{g_j} (\nabla \theta  _j)^2
	\right] .
\label{lutham}
\end{equation}
It is written as a sum of the contributions from the  spin ($j = \sigma $) and
charge ($j = \rho $) degrees of freedom. The parameters $g_j$ denote the
interaction strengths and $v_j=2v_F/g_j$  the velocities of spin and charge
excitations~\cite{Kane}. The parameters $g_j$ can be determined once one
defines
an appriopriate microscopic Hamiltonian (e.g. the Hubbard model); an
approximate
form for the spinless case has
 been given in  Refs.~\onlinecite{Kane,MatveevGlazman}. In this paper we will
neglect backscattering (via Umklapp or impurity scattering) and restrict
ourselves to repulsive, spin-independent interactions. As a result we have
$g_{\sigma} =2$ and $v_{\sigma} =v_F$~\cite{Kane}.

We also introduced bosonic fields
$\phi _{j}$ and $\theta _{j}$. They are related to the fields
$\phi _{s}= \phi _{\rho} +s \phi _{\sigma}$ and
$\theta _{s}= \theta _{\rho} +s \theta_{\sigma}$ for spin up ($s=+1$) and down
($s=-1$) fermions. These fields obey the commutation relation
$[\phi _s (x), \theta _{s'} (x')] =  (i \pi /2) \mbox{ sign} (x'-x) \delta
_{s,s'}$. The fermionic field operators $\hat{\Psi}$ can be expressed in terms
of the spin and charge degrees of freedom ~\cite{Loss,Haldane}:
\begin{eqnarray}
\hat{\Psi} ^{\dagger}_{L,s}
	(x)
&& =
	\sqrt{\rho _{0}}
	\sum \limits _{n, odd}
	\exp \left\{i n k_F x \right\} \nonumber\\
&& \times
	\exp \left\{i n [\theta _{\rho} +s \theta _{\sigma}]\right\}
	\exp \left\{i[\phi _{\rho} +s \phi _{\sigma}]\right\},
\label{fieldoperator}
\end{eqnarray}
where $k_F$ is the Fermi-wave vector and $\rho _{0} \equiv N_0/L $ is the
average electron density per spin direction. The number $N_0$ determines the
linearization point of the original electron spectrum,
$k_F \equiv \pi N_0 /L $~\cite{Loss}.

If the wire is closed to form a loop, the periodic condition~\cite{Loss}
\begin{equation}
\hat{\Psi}_{L,s}(x+L) = (-1)^{N_0} \hat{\Psi}_{L,s}(x).
\label{bc}
\end{equation}
should be imposed on the Fermi operators~(\ref{fieldoperator}). The fields
$\theta $ and $\phi $ can then be decomposed  in terms of bosonic fields
$\bar{\theta }$ and $\bar{\phi }$ and topological
excitations~\cite{Loss,Haldane}:
\begin{mathletters}
\begin{eqnarray}
	\theta _j(x)
	& = &
	\bar{\theta }_j(x) + \theta _j^0 + \pi M_j(x/2L) , \label{theta} \\
	\phi _j(x)
	& = &
	\bar{\phi } _j(x) + \phi _j^0 + \pi J_j [(x+L/2)/2L]
	.
\label{phi}
\end{eqnarray}
\end{mathletters}
Here, $\bar{\theta }_j$ and $\bar{\phi }_j$ are given by
\begin{mathletters}
\begin{eqnarray}
	\bar{\theta} _j (x)
	&=&
	\frac{i}{2} \sqrt{\frac{g_j}{2}} \sum _{q \ne 0}
	\left|
	\frac{\pi}{qL}
	\right|^{1/2}
	\mbox{sign}(q) e^{iqx}
	(\hat{b}^{\dagger}_{j,q}
	+
	\hat{b}_{j,-q}) ,\label{fieldsa} \\
	\bar{\phi} _j(x)
	&=&
	\frac{i}{2} \sqrt{\frac{2}{g_j}} \sum _{q \ne 0}
	\left|
	\frac{\pi}{qL}
	\right|^{1/2}
	e^{iqx}
	(\hat{b}^{\dagger}_{j,q}
	-
	 \hat{b}_{j,-q}) ,
\label{fieldsb}
\end{eqnarray}
\end{mathletters}
where $\hat{b}_{j,q}, \hat{b}^{\dagger}_{j,q}$ are Bose operators.

The boundary condition (\ref{bc}) gives  rise to the topological excitations
$M_j$ and $J_j$ for  the spin and charge degrees of freedom.  They are related
to the usual topological excitations for fermions with spin $s$: $M_s =
(1/2)[M_{\rho} + sM_{\sigma}]$ and
$J_s = (1/2)[J_{\rho} + sJ_{\sigma}]$. Physically, the number $M_s$ denotes the
number of excess electrons in the Luttinger liquid (LL) with spin $s$  in
addition to $N_0$. The number $J_s$ is the number of current quanta
$ev_F/L$, carried by electrons with spin $s$.
Here, $v_F = \pi \rho_0 /m$ is the Fermi velocity, with $m$ being the electron
mass. A net current $ J_\rho ev_F/L$
flows through the quantum wire if there is an imbalance between the number
electrons moving to the right and to the left. Using the boundary condition
(\ref{bc}) one obtains topological constraints for $M_s$ and $J_s$
\cite{Loss,Haldane},  which lead to the following constraints for $M_j$ and
$J_j$:\\ (i) The topological numbers
$M_j$ and $J_j$ are either {\em simultaneously} even or {\em simultaneously}
odd;\\  (ii) when $N_0$ is {\em odd} the sum
$M_{\rho} \pm M_{\sigma} + J_{\rho} \pm J_{\sigma}$ takes values $..., -4, 0,
4,
...$, when $N_0$ is {\em even} the sum
$M_{\rho} \pm M_{\sigma} + J_{\rho} \pm J_{\sigma}$ takes values $...,-6, -2,
2,
6, ...$.

An Aharonov-Bohm flux $\Phi$ threading the loop couples
to the net current, characterized by the  topological number $J_{\rho}$ of the
field $ \phi_{\rho}$.  The flux can be incorporated  into the Hamiltonian
(\ref{lutham}) by changing
\begin{equation}
\nabla \phi _j \rightarrow
\nabla \phi _j - (2\pi/L)  \delta_{j,\rho} f_{\Phi},
\label{changephi}
\end{equation}
where  $f_{\Phi} =  \Phi/ \Phi _0$ is the flux frustration and
$\Phi _0$ is the flux quantum $h/e$.

Since the LL is brought into the contact with  particle reservoirs
(superconductors) kept at fixed electro-chemical  potential $\mu$, the number
of
particles (characterized by the  topological number $M_{\rho}$ of the field
$\theta_{\rho}$)  should be coupled to $\mu$.  This can be achieved by
replacing
\begin{equation}
\nabla \theta _j \rightarrow
\nabla \theta _j - (2\pi/L)  \delta_{j,\rho} f_{\mu}
\label{changetheta}
\end{equation}
in the Hamiltonian (\ref{lutham}).  The parameter
$f_{\mu} =(g_{\rho}L/4\pi v_{\rho})\Delta \mu$  is related to the the
difference
$\Delta \mu$ between $\mu$ and the Fermi-energy $E_F =  k_F^2/2m$ of the
quantum
wire, corresponding to the linearization point. Generally, the
reference point
$\Delta \mu = 0$ is defined from the requirement that for $\Phi =0$ there are
$2N_0$ electrons in the ground state and the energies to add/remove electrons
to/from the system are equal. The difference $\Delta \mu$ can be controlled,
e.g., by a gate voltage.

Using Eqs.~(\ref{lutham}), (\ref{theta}), (\ref{phi}), (\ref{fieldsa}), and
(\ref{fieldsb}) one concludes that the Hamiltonian can be decomposed into
non-zero modes and topological excitations:
\begin{eqnarray}
&&	\displaystyle{\hat{H}_L
	=
	\sum _{j=\rho ,\sigma}
	\left\{
	\sum _{q\ne 0}
	v_j |q| \hat{b}^{\dagger}_{q,j} \hat{b}_{q,j} \right. }\nonumber \\
&&\displaystyle{\left. 	+
	\frac{\pi v_j}{4 L}
		\left[
		\frac{g_j}{2}
		(J_j - 4 \delta _{j,\rho} f_{\Phi}) ^2
		+
		\frac{2}{g_j} (M_j- 4 \delta _{j,\rho} f_{\mu})^2
		\right]
	\right\}} .
\label{luttop}
\end{eqnarray}
Since this Hamiltonian is quadratic in the Bose operators, it
is possible to obtain all the correlation functions exactly.

\section{DC-Josephson Effect}
Both systems depicted in Fig.~\ref{system} can be described by the Hamiltonian
\begin{equation}
        \hat{H}
        =
        \hat{H}_{S1} + \hat{H}_{S2} + \hat{H}_L + \hat{H}_T
        \equiv \hat{H}_0 + \hat{H}_T.
\label{sysham}
\end{equation}
Here, $\hat{H}_{S1}$, $\hat{H}_{S2}$ are the BCS-Hamiltonians for the bulk
superconductors kept at constant phase difference $\chi = \chi _{S1} -\chi
_{S2}$. For simplicity we assume equal magnitude $\Delta$ for both order
parameters. The tunneling between the superconductors and the 1D electron
system
is described by $\hat{H}_T$. It is assumed to occur through two  tunnel
junctions~\cite{footnote3} at the points $x=0$ and $x=d$,
\begin{eqnarray}
	\displaystyle{\hat{H}_T
	=
	\sum _{s}
	T _1 \hat{\Psi} ^{\dagger}_{S1,s} (x=0) \hat{\Psi}_{L,s}(x=0)} \nonumber \\
 \displaystyle{
	+
	T _2 \hat{\Psi} ^{\dagger}_{S2,s} (x=d) \hat{\Psi}_{L,s}(x=d)
	+
	\mbox{(h.c.)}}.
\label{tunnelham}
\end{eqnarray}
The constant tunnel matrix elements $T _{1,2}$  can be related
to  the tunnel conductances $G_{1,2}$ of the junctions,
$G_i = 4\pi e^2  N_L(0) N_i (0) |T_i|^2 $,
 where $N_L(0)= 1/\pi v_F$, and $N_i$ is the normal density of states in the
superconductors $(i=1,2)$.

The  stationary Josephson effect can be obtained by evaluating the
phase-dependent part of the free energy $\cal{F}(\chi)$.
The Josephson current is then given by
\begin{equation}
	I_J
	=
	- 2e \partial \cal{F} / \partial \chi .
\label{Ijdefinition}
\end{equation}
We expand ${\cal F} = -(1 / \beta ) \ln Z$, where
$Z = \mbox{Tr } \exp \{-\beta \hat{H}\}$ and $\beta = 1/ T$, in powers of the
tunneling Hamiltonian $\hat{H}_T$, using  standard imaginary-time perturbation
theory~\cite{Abrikosov}. The lowest order phase-dependent contribution arises
in
$4^{th}$ order. Using Eq.~(\ref{tunnelham}) we see that there are 24
contributions to the phase-dependent part of ${\cal F}$:
\begin{eqnarray}
\displaystyle{{\cal F}(\chi)
	=
	-
	\frac{1}{\beta}
 \int \limits _{0} ^{\beta} d\tau _1
 \int \limits _{0} ^{\tau _1} d\tau _2
 \int \limits _{0} ^{\tau _2} d\tau _3
 \int \limits _{0} ^{\tau _3} d\tau _4 } \nonumber \\
\displaystyle{ \times
 \left\{
	F_{S1}^{\dagger}(0; \tau _1 - \tau _2)
	T_1^2
	\Pi _L^{(a)}(0,d;\tau _1, ...,\tau _4) \right.} \nonumber \\
\displaystyle{ \left. \times
(T_2^*)^2
	F_{S2}(0; \tau_3 - \tau _4)
 \right. } \nonumber \\
\displaystyle{ \left. +
	F_{S1}^{\dagger}(0; \tau _1 - \tau _4)
	T_1^2
	\Pi _L^{(b)}(0,d;\tau _1, ...,\tau _4) \right.} \nonumber \\
\displaystyle{ \left.
\times
(T_2^*)^2
	F_{S2}(0; \tau_2 - \tau _3)
	+
	(\mbox{h.c.}) \right\} } \nonumber \\
\displaystyle{
+ 22 \mbox{ similar terms}
} .
\label{freeenergy}
\end{eqnarray}
This result has a clear physical meaning,  see Fig.~\ref{diagram1}. The
Josephson effect consists of processes in which a Cooper pair tunnels from
superconductor S2 into the LL with an amplitude $(T_2^*) ^2$. After propagation
through the LL, it tunnels into superconductor S1 with an amplitude $(T_1) ^2$.
The hermitian conjugate terms describe processes in the opposite direction. The
propagation in the superconductors is described by the anomalous Green's
function
\begin{eqnarray}
\displaystyle{
	F_{Si}(0;\tau -\tau ')
	\equiv
 \left\langle
	\mbox{T}_\tau
	\hat{\Psi}_{Si,-}(d,\tau)
	\hat{\Psi}_{Si,+}(d,\tau ')
 \right \rangle _{Si } }\nonumber \\
 \displaystyle{
 =
	\frac{\pi N(0)}{\beta}
	\sum _{\omega _n}  e^{-i\omega _n(\tau  - \tau ')}
 	\frac{\Delta e^{i\chi _{Si}}}{\sqrt{\omega _n^2 + \Delta^2}}},
\label{anomalous}
\end{eqnarray}
where $\langle ...\rangle _{Si}$ indicates an average with respect to $\hat
{H} _{Si}$.
Propagation through the LL is determined by the Cooperon propagator
$	\Pi _L(0,d;\tau _1, ...,\tau _4)$. These 24 terms are obtained by considering
all possible time-ordered pairs of tunneling events $\tau _i , \tau _j$ ($\tau
_i < \tau _j$) at
$x=0$ and $x=d$ together with all possible spin configurations. However, which
of these terms are important depends on the relation between the characteristic
energy $ v_F/d$ for the 1D system and the superconducting gap
$\Delta$~\cite{footnote5}.

If the distance between the contacts is large, $ v_F/d \ll \Delta$, a
generic process consists of fast tunneling of two electrons from the
superconductor into the 1D system and their slow propagation through the LL.
Such a process is illustrated by the first term in Eq.~(\ref{freeenergy}).
Here,
\begin{eqnarray}
&&\Pi _L^{(a)}(0,d;\tau _1, ...,\tau _4) \nonumber \\
&&\displaystyle{=
	\left\langle
	\hat{\Psi}_{L,+}(0,\tau _1)
	\hat{\Psi}_{L,-}(0,\tau _2)
	\hat{\Psi} ^{\dagger}_{L,+} (d,\tau _3)
	\hat{\Psi} ^{\dagger}_{L,-} (d,\tau _4)
	\right\rangle}  ,
\label{cooperonslow}
\end{eqnarray}
where
$|\tau_1-\tau_3| \sim d/v_F \gg |\tau_1-\tau_2| \sim
|\tau_3-\tau_4| \sim 1/\Delta$
(see Fig.~\ref{diagram1}a).
The average is taken over equilibrium fluctuations in the LL  (described by
$\hat{H}_L$).  The other relevant processes come from diagrams which are
obtained from the one in Fig.~\ref{diagram1}a by means of particle-hole
conjugation and by changing the time ordering.

In the opposite limit $ v_F/d \gg \Delta$, diagrams of the type depicted
in Fig.~\ref{diagram1}a are no longer relevant.  Instead, one should consider
fast and independent propagation  of two electrons through the LL and slow
tunneling between S and LL. This is illustrated by the second term in
Eq.~(\ref{freeenergy}) (see also Fig.~\ref{diagram1}b), where
\begin{eqnarray}
&&\Pi  _L^{(b)}(0,d;\tau _1, ...,\tau _4) \nonumber \\
&&\displaystyle{ =
	\left\langle
	\hat{\Psi} _{L,+}(0,\tau _1)
	\hat{\Psi}^{\dagger}_{L,+}(d,\tau _2)
	\hat{\Psi}_{L,-} (0,\tau _3)
	\hat{\Psi}^{\dagger} _{L,-} (d,\tau _4)
	\right\rangle},
\label{cooperonfast}
\end{eqnarray}
with
$|\tau_1-\tau_2| \sim  |\tau_3-\tau_4| \sim d/v_F \ll |\tau_1-\tau_3| \sim
1/\Delta$. Also in this case the other relevant processes  can be obtained from
the one in Fig.~\ref{diagram1}b by means of particle-hole conjugation and by
changing the time ordering.

The direct evaluation of averages  like~(\ref{cooperonslow}),
(\ref{cooperonfast}) with the help of  bosonized field operators
like~(\ref{fieldoperator}) is tedious but  straightforward. The resulting
expressions  can be simplified further in the two limiting cases
$ v_F/d \ll \Delta$ and $ v_F/d \gg \Delta$, which contain all the important
physics of the problem.

\section{DC-Josephson Current through a
Quantum Wire}
We first turn to the geometry depicted in Fig.~\ref{system}a. It consists of a
quantum wire of length $L \to \infty$ connected to two superconductors by
tunnel
junctions separated by a distance $ d$. The topological excitations play no
role
in this case  (their energy is vanishingly small) and  the wire is described by
the non-zero modes only (first term in Eq.(\ref{luttop})).\\

{\em a. The case $ v_F/d \ll \Delta$.}

The expression for the phase-dependent part of the free energy contains four
terms of the type of the first term in Eq.~(\ref{freeenergy}) (see
Fig.~\ref{diagram1}a). In this case at low temperatures  $ T \ll \Delta$ one
can  approximate the anomalous Green functions (\ref{anomalous}) by
$\delta$-functions in time. This fixes equal time arguments $\tau_1=\tau_2$,
$\tau_3=\tau_4$ in $\Pi _{L}^{(a)}$. The remaining integration should be
performed over the time $\tau =
\tau_1 - \tau_3$. The dominant contribution to the the integral comes from  the
terms with $n= \pm 1$ in Eq. (\ref{fieldoperator}). As a result, the Josephson
current (\ref{Ijdefinition}) through the quantum wire is given by
\begin{equation}
I_J ^{(a)} = I_c^{(a)} (T) \sin \chi,
\label{ijt}
\end{equation}
with a temperature-dependent critical current
\begin{equation}
	I_c^{(a)} (T)
	=
	\frac{4\pi e v_F}{d} \frac{G_1 G_2}{(4e^2)^2}
	F_{w}^{(a)}(T),
\label{icrt}
\end{equation}
where
\begin{eqnarray}
\displaystyle{
	F_{w}^{(a)}(T)
	=
	\left[\frac{1}{k_Fd}\right]^{2/g_{\rho} - 1}
 \int \limits _{\;\:-\frac{\beta v_F}{2d}}
                     ^{\;\frac{\beta v_F}{2d}}
	\frac{dx}{2\pi}\prod_{j=\rho,\sigma}
	\left[\frac{2\pi ^2 d^2}{v_j^2\beta ^2} \right. }&& \nonumber \\
\displaystyle{
	\times
	\left.
 	\frac{2}{\cosh\left(2\pi d/v_j\beta \right) -
	        \cos\left(2\pi d x/\beta v_F \right)}
	\right]^{\frac{1}{g_j}} }&&
\label{piappr}
\end{eqnarray}
(with $g_{\sigma}=2$ and $v_{\sigma} = v_F$).

In the non-interacting case, at zero temperature,
$F_w^{(a)}(0) = 1$.
The Josephson current decreases  as $1/d$ with
increasing distance between the tunnel junctions.
This is related to the fact that
the density of Cooper pairs in the LL decays in space
away from each junction. Hence the overlap of the macroscopic wave functions of
the two superconductors, which is responsible for the Josephson effect, is
suppressed. Repulsive interactions in the wire make the Josephson effect vanish
more rapidly with the distance between the superconductors,
$$
I_c ^{(a)}(0)\propto 1/(k_Fd)^{2/g_{\rho}}\:\:.
$$
The electron liquid acquires an additional stiffness against density
fluctuations, hence the tunneling between S and LL is suppressed. This fact
provides an {\em a posteriori} justification of our use of perturbation theory
when treating electron tunneling in the presence of repulsive
interactions.

We consider now the temperature dependence of the critical current. For
non-interacting electrons, $g_{\rho} =2$, the critical current can be
calculated
explicitly:
\begin{equation}
\frac{I_{c}^{(a)}(T)}{I_{c}^{(a)}(0)} =
\frac{2\pi T d}{ v_F}
	\frac{1}
{\sqrt{\cosh^2(2\pi T d/ v_F)-1}} .
\end{equation}
At low temperatures, $T \ll  v_F/2\pi d$,  the critical current is
suppressed below its zero temperature value
in a power law fashion
\begin{equation}
I_{c}^{(a)}(T) / I_{c}^{(a)}(0)
\simeq 1 - \frac{2}{3} \left( \frac{\pi T d}{ v_F} \right)^2 \: \: .
\label{lowT}
\end{equation}
In the high temperature regime,
$T \ll  v_F/2\pi d$, the decay is exponential
$$
\frac{I_{c}^{(a)}(T)}{I_{c}^{(a)}(0)} \simeq
	\frac{\sqrt{8} \pi T d}{ v_F }
	\exp(-2\pi T d/ v_F ).
$$

It is possible to obtain
analytical results also in the interacting case.
In particular, for weak interaction $2-g_{\rho} \ll 2$ and
low temperatures $T \ll  v_F/2\pi d$
the critical current behaves as
\begin{equation}
\frac{I_{c}^{(a)}(T)}{I_{c}^{(a)}(0)}
\simeq 1 + \frac{2-g_{\rho}}{4}
\left( \frac{2 \pi T d}{v_F} \right)^{2/g_{\rho}}
- \frac{2}{3}\left(\frac{\pi Td}{v_F}\right)^2 ,
\label{anomalicr}
\end{equation}
where we dropped terms
$ {\cal O}[ (2-g_{\rho})^2 + (2-g_{\rho}) (Td/v_F)^2 ]$.
This result can be interpreted in terms of a competition between two effects.
At low temperatures the dominant dependence comes from the renormalization of
the tunneling amplitudes $T_{1,2}$ in the presence of interaction~\cite{Kane}.
The critical current {\em increases} with temperature. Above the crossover
temperature $T_{\rm cross} \simeq (3/8) (2-g_{\rho}) v_F/ \pi d$ the Josephson
current  {\em decreases}  due to the shortening of the phase coherence length.
Although the maximum is not very pronounced, the cross-over temperature shifts
to  higher values as the interaction strength increases  (see
Fig.~\ref{iwiretemp}). This results in a wider temperature range in which  the
critical stays almost constant.  It is evident from  Eq.~(\ref{anomalicr}) that
the coefficient  responsible for the anomalous temperature dependence vanishes
in the  absence of interaction, thus restoring the $T^2$ suppression of the
critical current  (\ref{lowT}). For high temperatures
$T \gg  v_{\rho}/2\pi d$ the suppression becomes exponential,
\begin{equation}
I_c^{(a)} (T)
\propto
T^{2/g_{\rho}} \exp(-2\pi T d/v_F) .
\end{equation}

The full temperature dependence of the critical current, calculated by
numerical
integration of Eqs.~(\ref{icrt}), (\ref{piappr}) is shown in
Fig.~\ref{iwiretemp}.  We see that for moderate strength of the interaction
$g \sim 1$ the Josephson current  will maintain an appreciable value up to a
temperature $T \sim  v_F / d$, which is of the order of $0.7$ K  for typical
experimental  parameters $v_F =$ 3.0 10$^5$ m/s and $d=3$ $\mu$m~\cite{Mailly}.
Moreover, the value of the
critical current $I_{c} ^{(a)}(T=0) \approx $ 22 nA, (estimated for the
parameters given above and $G_i/(4e^2) = 0.3$) is large enough to be
measured experimentally.

Note that we estimated the Josephson current assuming fixed Josephson phase
difference between the  superconductors. Thermal fluctuations of the Josephson
phase would smear the critical current at temperatures $T^* \sim E_J \equiv
I_c
^{(a)}/2e$, provided that the superconductors are coupled by the LL only. Using
Eq.~(\ref{icrt}) for the non-interacting case
(with $F_w^{(a)}(0) = 1$), one  obtains that the temperature $T^*$ is
by a factor $2\pi G_1 G_2 /(4e^2)^2 \ll 1$ smaller than the characteristic
temperature scale $v_F/d$ for the LL .
Hence, in order to observe non-trivial temperature dependence of the critical
current, one has to fix the phase difference between the superconductors,
e.g., by means of an additional Josephson junction.\\

{\em b. The case $v_F/d \gg \Delta$.}

In this limit, the electrons propagate fast and independently through the LL on
a time scale $1/\Delta$. A typical contribution is depicted in
Fig.~\ref{diagram1}b. The Cooperon~(\ref{cooperonfast}) can be approximated as
\begin{eqnarray}
&&\Pi^{(b)}_L(0,d;\tau _1,...,\tau _4)
\approx \nonumber \\
&&	\left\langle
	\hat{\Psi} _{L,+}(0,\tau _1)
	\hat{\Psi}^{\dagger}_{L,+}(d,\tau _2)
\right\rangle
\left\langle
	\hat{\Psi}_{L,-} (0,\tau _3)
	\hat{\Psi}^{\dagger} _{L,-} (d,\tau _4)
\right\rangle, \nonumber
\end{eqnarray}
where we substitute
$$
	\left\langle
	\hat{\Psi} _{L,+}(0,\tau _i)
	\hat{\Psi}^{\dagger}_{L,+}(d,\tau _j)
\right\rangle
\approx
C \delta(\tau _i - \tau _j) .
$$
The constant $C$ is determined by integration of the time-ordered
single particle correlator of the LL,
$$
C = \int \limits _{-\beta/2} ^{\beta/2} d\tau
	\left\langle
 \mbox{T}_\tau
	\hat{\Psi} _{L,+}(0,\tau)
	\hat{\Psi}^{\dagger}_{L,+}(d,0)
\right\rangle .
$$
The temperature-dependent Josephson current is found to be
$$
I_J^{(b)} = I_c^{(b)}(T) \sin \chi ,
$$
where the critical current is given by
$$
	I_c^{(b)}(T)
	=
e\Delta \frac{G_1G_2}{(4e^2/\pi)^2} F_w ^{(b)},
$$
with
\begin{eqnarray}
&&\displaystyle{ F_w ^{(b)}
=\left[\frac{1}{k_Fd}\right]^{g_{\rho}/4 + 1/g_{\rho} -1}
\left[ \frac{g_\rho}{2}\right]^{g_{\rho}/4 + 1/g_{\rho}}
\left[ \frac{\pi d}{\beta v_F}\right]^{\frac{g_{\rho}}{4} +
\frac{1}{g_{\rho}}+1} }
\nonumber
\\ &&\displaystyle{ \times \left\{ \frac{2}{\pi}
\int \limits_{0} ^{\beta v_F/2d} dx
\sin\left(\frac{\zeta _{\rho} + \zeta _{\sigma}}{2}\right) \right.}\nonumber \\
&&\displaystyle{\times \left. \prod_{j=\rho,\sigma}
\left[\frac{2}{\cosh (2\pi d/\beta v_j) - \cos 2\pi d x/\beta
v_F}\right]^{\frac{g_j}{16} + \frac{1}{4g_j}}
\right\}^2}
\label{shortF} .
\end{eqnarray}
The phase factor $\zeta _j $ is given by
$$
\zeta _j = \arctan \left[ \cot \left(\frac{\pi d x}{\beta v_F}\right) \tanh
\left(\frac{\pi d}{\beta v_j}\right)\right].
$$

In the non-interacting case $g_\rho = g_{\sigma}=2$, $F_w^{(b)}$ can be
calculated explicitly:
$$
F_w^{(b)}
=
\left[\frac{2}{\pi}
\arctan\left(\frac{1}{\sinh(\pi d /\beta v_F)}\right)
\right]^2 .
$$
At zero temperature we find $F_w^{(b)}=1$.
The resulting Josephson current thus is independent of the
distance $d$ between the contacts,
analogous to the result obtained in Ref.~\onlinecite{Beenakkerquant}.
At finite temperatures the Josephson current is suppressed.
If $
T \ll  v_F/d$, the suppression is linear in $T$:
$$
\frac{I_c^{(b)}(T)}{I_c^{(b)}(0)} \simeq 1 -  \frac{4Td}{ v_F }.
$$

In the interacting case, at $T=0$, the critical current is suppressed,
$$
I_c^{(b)}(0) \propto 1/(k_Fd)^{g_{\rho}/4 + 1/g_{\rho} -1}.
$$
At finite temperatures, and for weak interactions we obtain
\begin{equation}
\frac{I_c^{(b)}(T)}{I_c^{(b)}(0)} -1  \sim
 -
\frac{3}{2\pi}
\left[\frac{\pi Td }{ v_F}\right]^{1/2 + g_{\rho}/8 +1/2g_{\rho}}
+ \frac{2-g_{\rho}}{2}\frac{\pi Td}{ v_F} ,
\end{equation}
where we dropped terms of the order of
$ {\cal O}[ (2-g_{\rho})^2 + (2-g_{\rho}) (Td/v_F)^2 ]$.
We find again an anomalous dependence on temperature, like the one we discussed
in the case $v_F/ d \ll \Delta$.

\section{DC-Josephson Current through a Ring}
In case of the ring with circumference $L$ (Fig.~\ref{system}b),  we should
take
into account the contribution to the Josephson current  due to the topological
part, see Eqs.~(\ref{theta}), (\ref{phi}), and (\ref{luttop}). The Cooperon for
the ring in case of a symmetric setup $d = L/2 \gg  v_F /\Delta$ is evaluated
along the same lines as before.  It is then straightforward to get the
Josephson
current
\begin{eqnarray}
\displaystyle{
	I_J
	=
	\frac{2 \pi e v_F}{L}\frac{G_1 G_2}{(4e^2)^2}
	\sum _{\epsilon = \pm 1}
 	\left\langle
	F_{r}(g_{\rho},L,\epsilon, M_{\sigma}, J_{\rho}) \right.}\nonumber \\
\displaystyle{ \left.
\times
	\sin{(\chi + \epsilon \pi M_{\sigma}/2 + \pi J_{\rho}/2)}
	\right\rangle _{J,M} },
\label{josring}
\end{eqnarray}
where
\begin{eqnarray}
&&\displaystyle{
	F_{r}
	=
 \frac{1}{2}
	\left[
	\frac{\pi}{k_FL}
	\right]^{2/g_{\rho} - 1}
	\int \limits _{-\pi \beta v_F/2L} ^{\pi \beta v_F/2L}
 dx
 \left[
	\frac{1}{\cosh(2x/g_{\rho})}
	\right]^{\frac{2}{g_{\rho}}} } \nonumber \\
&&\displaystyle{
	\times \frac{1}{\cosh(x)}
 \cosh
	\left[
	\left(\frac{2}{g_{\rho}}\right)^2
	(M_\rho - 4f_{\mu}) x
	+
	\epsilon
	J_\sigma x
	\right] } \nonumber \\
&&\displaystyle{ \times
\prod _{j=\rho, \sigma}
		\left[
		\frac{1+\sum_{n}\exp\left(2\pi \beta v_j
		n^2/L\right)}{1+\sum_{n}\exp\left(2\pi \beta v_j
		n^2/L\right) \cos(2n \lambda_j(x))}
		\right]^{\frac{1}{g _j}}},
\label{fring}
\end{eqnarray}
with
$$
\lambda_j(x) = \frac{1}{2}
\mbox{arcos}\left[\cosh\left(\frac{2v_j x}{ v_F}\right)\right] .
$$
The brackets, $\langle ... \rangle _{J,M}$, should be considered  as the
thermal
average over the topological excitations weighted by the appropriate Boltzmann
factor and subject to the topological constraints.

For zero temperature this calculation involves only the ground state (see
Ref.~\onlinecite{Fazio} for details). The dependence of the critical current on
$f_{\Phi}$ and $f_{\mu}$ has been found to show very rich behavior. In the
present study, we will focus on the effect of finite temperatures.  In
particular, we will investigate how robust the structure, found in
Ref.~\onlinecite{Fazio} is against thermal fluctuations.

Two remarks are in order at this point. First, throughout this section, we
assume that the linearization point of the original electron
spectrum corresponds to odd values of $N_0$ (for even $N_0$ the picture is the
same, apart from a relative shift of $f_{\Phi}$ and $f_{\mu}$). Second, the
Josephson current depends periodically on both $f_{\Phi}$ and $f_{\mu}$ with
period 1. However, since the original problem has additional symmetries
$f_{\Phi} \rightarrow -f_{\Phi}$ and $f_{\mu}
\rightarrow -f_{\mu}$  (together with a change of sign of the corresponding
topological numbers, $J_j$ and $M_j$), it is enough to consider $f_{\Phi}$ and
$f_{\mu}$ in the intervals $0<f_{\Phi}<1/2$ and $0<f_{\mu}<1/2$.

It is instructive to discuss first the non-interacting case $g_{\rho } = 2$,
for
$T=0$. If $f_\mu +f_\Phi < 1/2$ the ground-state configuration of $J_j$ and
$M_j$ is found to be
$(J_{\rho},J_{\sigma},M_{\rho},M_{\sigma}) =$ (0,0,0,0); if $ f_\mu +f_\Phi >
1/2$ the configuration is (2,0,2,0). Hence, the ground-state of the system can
be changed by varying  either the flux or the gate voltage. As a result, the
Josephson current changes, see Eqs.~(\ref{josring}), (\ref{fring}). This is
illustrated in Fig.~\ref{iringmug2}, where the critical current
$I_c$ (we write $I_J = I_c \sin \chi$ where $I_c$ can be positive or negative)
is plotted as a function of $f_{\mu}$ and $T$ at fixed $f_{\Phi} = 0.2$. For
$T=0$,  the critical current shows a maximum and a sharp jump at $f_{\mu} =
0.3$
where the states (0,0,0,0) and (2,0,2,0) are degenerate. At this value of the
gate voltage, the number of electrons on the ring ($M_{\rho}$)  increases by
two.
Since electronic states are doubly degenerate in spin  and non-zero flux is
applied, the two electrons will occupy the same  (clockwise or counterclockwise
moving) single-particle state. Therefore, the net current $ev_FJ_{\rho}/L$
increases by 2 quanta
$ev_F/L$ while the topological numbers $M_{\sigma}$ and $J_{\sigma}$ related to
spin remain unchanged. At the jump, $I_c$ changes sign. This reflects the fact
that the ring acts as a $\pi$-junction  ($I_c < 0$) in the state (2,0,2,0), as
can be seen from Eq.~(\ref{josring}). Therefore, for non-interacting electrons,
the critical current shows two jumps per period of the gate voltage dependence.
The same is true for the dependence of $I_c$ on the magnetic flux.

This picture is correct for any generic point on the line
$ f_\mu +f_\Phi = 1/2$. At the end points
$(f_{\mu}, f_\Phi) = (0,0.5)$ and $(0.5,0)$,  no jumps of the critical current
occur (one can say that two jumps  in opposite directions merge together).
Instead, the critical current  shows a resonance. The resonance occurs due to
alignment of two spin degenerate energy levels (for clockwise and
counterclockwise moving electrons) with the  chemical potential of the
superconductors~\cite{Fazio}.

At a finite temperature, both the non-zero modes and the topological
excitations
are thermally activated.  Thermal activation of the non-zero modes leads to an
overall suppression of the critical current, as it has been discussed  for  the
wire in Section 4.  Thermal activation of the states (0,0,0,0) and (2,0,2,0)
will
lead to a smearing of the jump. Moreover, at finite temperature there will be a
non-vanishing probability to activate other topological excitations which can
contribute to the Josephson current. In the plotted temperature range, only one
additional state (1,1,1,1) with one extra electron on the ring can be
activated.  As a result, the negative  critical current of the state (2,0,2,0)
(at $f_{\mu} \agt 0.3$) will be partially compensated  by the positive critical
current of the state (1,1,1,1), the occupancy of which increases with
temperature. Note that the jump at $f_\mu =0.3$ remains visible up to
temperatures of the order of $T
\sim  v_F/ L
\sim 1$K (for the parameters mentioned above and $L = 2 \mu$m). Hence, the
parity
effect causing the jump is quite robust against thermal fluctuations.

An important feature of the non-interacting case at $T=0$ is that the various
possible ground-state configurations may differ by an {\em even} number of
electrons only. The situation changes  drastically when repulsive interactions
are switched on. In addition to the states (0,0,0,0) and (2,0,2,0), the state
(1,1,1,1) can act as a ground state configuration~\cite{Fazio}. This happens
for
parameters $f_\mu$, $f_\Phi$ within the range
$1+3(g_\rho /2)^2  < 8[f_\mu + (g_\rho/2)^2 f_\Phi] < 3+(g_\rho /2)^2 $. Within
this "strip", it is energetically more favorable to add a {\em single}
electron,
rather than a pair of electrons to the ring, due to repulsive electron-electron
interactions. The Josephson current  in the state (1,1,1,1)  differs from the
current  in the states (0,0,0,0) and (2,0,2,0), see Eqs.~(\ref{josring}),
(\ref{fring}). For example, for $g_{\rho} = 1.75$ and $f_{\Phi} = 0.2$
(Fig.~\ref{iringmug175})  the state (1,1,1,1) occurs in the range
$0.259 < f_{\mu} < 0.318$. Indeed, one sees two pronounced jumps of
$I_c$ at the borders of this interval in Fig.~\ref{iringmug175}  (at low
temperatures). Generally, for interacting electrons   the critical current
shows
four jumps per period of the gate voltage dependence \cite{Fazio}.

Similar jumps are seen also at the dependence of the critical current on the
flux, Fig.~\ref{iringfluxg175}. Moreover, $I_c$ is a stepwise function of
$f_{\phi}$ for $T=0$ (this can be seen from Eqs.~(\ref{josring}),
(\ref{fring});
the flux $f_\Phi$ enters to these equations only implicitly, via topological
numbers). Depending on the gate voltage, the critical current can show zero,
two, or four jumps per period of the flux dependence \cite{Fazio}.

The state (1,1,1,1) is the ground state in a strip of width $\delta f_\mu
= [1-(g_\rho/2)^2]/4$. This determines the energy
$\delta E \simeq  [1-(g_\rho/2)^2] \pi v_{\rho}/(g_{\rho} L)$  needed to create
topological excitations. The features related to the state (1,1,1,1) will be
smeared at temperatures $  T \sim \delta E$. Therefore, for weak interaction
$2-g_{\rho} \ll 1$, the interaction effects will dissapear at much lower
temperatures $  T \ll  v_F/L$  than the parity effects. For example,  the
features related to the configuration (1,1,1,1) in  Figs.~\ref{iringmug175},
\ref{iringfluxg175} are seen only in the temperature range $T < \delta E  \sim
0.1$ K whereas the overall  dependence is robust up to the temperatures $T \sim
1$K. However, it is worthwhile to stress that the (1,1,1,1) state survives at
much higher temperatures when the interaction strength is increased.

The behavior we described here is rather generic for all values of $f_\mu$,
$f_\Phi$  and for various values of the interaction. What is specific is
the configuration of the two superconductors: they are connected symmetrically
to the ring. If the points on the ring at which the  electrodes are attached
would form a generic angle, a more complicated interference pattern would
arise.
In the symmetic setup, the maximum Josephson current occurs either  at $\chi
=0$
or at $\chi =\pi$.  In the nonsymmetric setup the maximum Josephson current
would occur at  a value
$\chi(J_{\rho},J_{\sigma},M_{\rho},M_{\sigma})$ which depends on the values of
topological numbers. The critical current should  then be found by maximizing
the resulting function of the phase difference.

\section{AC-Josephson Effect}

The effect of a finite DC bias voltage $eV \ll 2\Delta$  applied between the
superconductors S1 and S2,  will be twofold. First of all, the phase difference
$\chi$ between S1 and S2 will acquire
a time-dependence, according to the Josephson relation
$\dot {\chi} =
\omega_J = 2eV $.
As a result, the Josephson current will oscillate as a function of time at a
frequency $\omega _J
$
(AC-Josephson effect, see Ref.~\onlinecite{Tinkham}). Secondly, a DC subgap
current will be induced, due to Andreev reflection at both junctions. This
current is dissipative, energy will be dissipated in the LL. In a typical
experiment one thus will find a current with both a DC and an AC component. In
this section, we will mainly concentrate on the AC Josephson current, and
estimate the DC component at the end.

In the presence of a bias voltage $V$ between the superconductors,
the imaginary time formalism cannot be applied and
Josephson current is  found by calculating  the average of the corresponding
Heisenberg operator.
Using the interaction representation with the unperturbed Hamiltonian
$\hat{H}_0$, see Eq.~(\ref{sysham}),
one obtains
\[
I(t) =  \langle \hat{U}^\dagger(t) \hat{I} (t) \hat{U}(t)  \rangle,
\]
\begin{equation}
\hat{U}(t) = T \exp (- i \int_{- \infty}^{t} \hat{H}_T (t') dt' ).
\label{ij(t)}
\end{equation}
We expand (\ref{ij(t)}) to the fourth order in $\hat{H}_T$ and keep the
Josephson terms in the current. These are proportional to $\exp(\pm 2ieVt)$.
As a result the Josephson current is given by
an expression which has the same structure as Eq.~(\ref{freeenergy}).
The integrals are now taken over real times. It is convenient to
depict the times $t, t_1, t_2, t_3$ of tunneling events on the Keldysh
contour~\cite{Keldish} (the Josephson current is calculated at a time $t$).
Again, we will consider two cases of long ($ v_F/d \ll \Delta$)
and short ($ v_F/d  \gg \Delta$) distance between the contacts.
The relevant diagrams are shown in Fig.~\ref{keldiagr1} for both cases.
We restrict our consideration to the case of a quantum wire at zero
temperature.\\

{\em a. The case $ v_F/d \ll \Delta$.}

For a large distance between the contacts
the tunneling of two electrons to/from a superconductor
is a fast process on the time scale of their propagation through LL.
The Josephson current is described by diagrams of the type
shown in Fig.~\ref{keldiagr1}a.
The Josephson current is then given by
\begin{eqnarray}
&\displaystyle{
I_J(t) = 4\pi ^2ev_F^2\frac{G_1 G_2}{(4e^2)^2 }}& \nonumber \\
&\displaystyle{ \times
\Re\mbox{e}
\left[
\sum_{\pm} \pm e^{\pm 2ieVt}
\int_0^{\infty} dt' e^{\mp ieVt'} \Pi(t')
\right]},&
\end{eqnarray}
where
$\Pi(t) = \Pi_L^{(a)}(0,d; it,it,0,0)$ is the Cooperon propagator
(\ref{cooperonslow})
in real time taken at coinciding
time arguments.  The leading contribution stems from the terms in
Eq.~(\ref{fieldoperator}) with $n =\pm 1$,
\begin{eqnarray}
&&
\Pi(t) = \nonumber \\
&&\displaystyle{2 \rho_0^2 \prod_{j=\rho,\sigma}
\{ [ 1 + i k_F (v_{j} t + d) ]
   [ 1 + i k_F (v_{j} t - d) ]\}^{-1/g_j}}.
\label{cooprealtime1}
\end{eqnarray}
In particular, for non-interacting electrons ($g_{\rho} =2$)
we obtain
\begin{equation}
I_J(t) = \frac{2 \pi e v_F}{d} \frac{G_1G_2}{(4e^2)^2}
\sin \left( 2eVt - \frac{eVd}{ v_F} \right).
\label{I_J(t)_nonint}
\end{equation}
This result means that the Josephson current acquires an additional
phase shift due to the propagation of electrons between the contacts.
For interacting electrons we computed the Josephson current numerically.
We split $I_J$ into sinusoidal and cosinusoidal
components,
\begin{eqnarray}
\displaystyle{
I_J(t) = \frac{2 \pi e v_F}{d} \frac{G_1G_2}{(4e^2)^2}
\left( \frac{1}{k_F d} \right)^{2/g_{\rho} -1}} \nonumber \\
\displaystyle{ \times
\left\{ J_s \left(\frac{eV}{ v_F/d}, g_{\rho} \right)
		  \sin \left( 2eVt \right) \right.}\nonumber \\
\displaystyle{\left.
     +  J_c \left(\frac{eV}{ v_F/d}, g_{\rho} \right)
		  \cos \left( 2eVt \right)
\right\}}.
\label{I_J(t)_gen}
\end{eqnarray}
The amplitudes $J_{c (s)}$ of the two components and the phase
$\varphi = -\arctan (J_c/J_s)$ of the Josephson current
are shown in Fig.~\ref{J_cs(V)} as functions of the voltage
for two values of the interaction parameter, $g_{\rho} = 1.75$ and $1$.
One sees that the deviation from the simple result (\ref{I_J(t)_nonint})
for non-interacting electrons
(which corresponds to
$ J_s =   \cos( eVd / v_F )$,
$ J_c = - \sin( eVd / v_F )$, and
$  \varphi =  eVd / v_F$)
increases with the increase of the interaction.
This deviation becomes striking in the dependence of
the AC current amplitude $J_a = \sqrt{J_s^2 + J_c^2}$ on
the voltage, Fig.~\ref{J_a(V)}. Apart from the non-interacting case
($J_a (V) = const$), one sees pronounced oscillations of
the current amplitude. These oscillations are due to the
difference in the velocities of the charge ($v_{\rho}$) and
spin ($v_{\sigma}$) excitations.
The period $\delta V$ of the oscillations corresponds to
$2\pi$ difference between the phases
of charge $(e V d/ v_{\rho})$
and spin  $(e V d/ v_F)$ excitations.
Using the relation $v_{\rho} = 2v_F / g_{\rho}$ we obtain
$e \delta V/( v_F/d) = 2 \pi (1-g_{\rho}/2)^{-1}
\simeq 50.4, 12.6$ for $g_{\rho} = 1.75$ and $1$, respectively. This is in very
good argeement with the period of oscillations in Fig.~\ref{J_a(V)}.
Therefore, the AC Josephson effect can be used as a tool for
the observation of {\em spin-charge} separation in the LL.\\

{\em b. The case $ v_F/d \gg \Delta$.}

At short distances between the contacts, the two electrons
propagate fast through the LL on the time scale $1/\Delta$.
The relevant diagrams are similar to the graph shown in
Fig.~\ref{keldiagr1}b.
The main contribution to the Josephson current comes from the integration
of the two-particle propagators of the type
$\Pi_{L}^{(b)} (0,d; it,it_1,it_2,it_3)$,
Eq. (\ref{cooperonfast})
(with possible permutations of creation and annihilation operators)
over the range
$t-t_1 \sim t_2-t_3 \sim d/v_F$ and $t-t_2 \sim 1/\Delta$
(see Fig.~\ref{keldiagr1}b).
For this reason, we can present $\Pi_L^{(b)}$ as a product of two
single-particle propagators and integrate the latter over the "fast"
variables $t-t_1$ and $t_2-t_3$ (from $0$ to $\infty$)
as we did for the DC case.
The last integration over the "slow" variable $s = t-t_2$ involves the
product of two anomalous Green functions with an exponent containing the
time-dependent Josephson phase:
$$
\int_0^{\infty} ds F_{S1}^{+}(0,is) F_{S2}(0,is) \exp( ieVs).
$$
Hence, for short distances between the contacts
the presence of LL does not
influence the voltage dependence of AC Josephson current.
The latter is still given by the simple formula
$$
I_J(t) = (2/\pi) K (eV/2\Delta) I_c ^{(b)}(0)\sin(2eVt) \:,
$$
where $K(x)$ is an elliptic integral and $I_c ^{(b)}(0)$ is
the critical current  in the DC case
(cf. Eq.~(\ref{shortF}) in the limit of zero temperature).
The effect of the interaction is
only to reduce the value of the critical current
while its voltage dependence is analogous to that of the critical current of a
conventional Josephson junction~\cite{Kulik}.\\

We conclude this section with an estimate of the dissipative DC current due to
Andreev reflection at both junctions. For a single junction between a
superconductor and a LL with repulsive interactions, the subgap current $I_s
(V)$
as a function of the applied voltage $V$ is given
by~\cite{Maslov,Fisher,Winkelholz}
$I_s (V)
\sim V|V|^{2/g_\rho -1}
$.
For the system of Fig.~1a, which consists of two junctions in series,
the lowest order contribution to the subgap current stems from
sequential tunneling. Employing a rate equation approach,
we find for this contribution
\begin{eqnarray}
\displaystyle{
I_s (V)
=
2\pi e \frac{G_1G_2}{(4e^2)^2} 2eV
\left[\frac{2eV}{ v_Fk_F} \right]^{2/g_\rho -1}
\frac{(2/g_\rho)^{2/g_\rho}}{\Gamma (1+2/g_\rho)} }\nonumber \\
\displaystyle{ \times
\left[
\frac{2(G_1G_2)^{g_\rho/2}}{(G_1^2)^{g_\rho/2} +(G_2^2)^{g_\rho/2}}
\right]^{2/g_\rho}}.
\end{eqnarray}
Comparing this result with the critical current we see that the dissipative
component is much smaller at low voltages. In order to get a complete
description at finite voltages, one has to solve
the corresponding equation for non-linear RSJ-model~\cite{Aslamasov}.
This will be discussed in a forthcoming publication~\cite{Faziorev}.

\section{Discussion}
In this paper, we studied the AC and DC Josephson effect in
a single mode quantum wire and quantum ring
connected to two superconductors by tunnel junctions.
Repulsive interactions were treated
in the framework of the Luttinger model.
Interactions were found to have a drastic influence on
both DC and AC Josephson effect.

The critical current is suppressed by
interactions at zero temperature.
The results depend on the ratio between the characteristic energy
$\hbar v_F/d$ of the 1D electron system and
the superconducting energy gap $\Delta$.
For large distances between the contacts $d \gg \hbar v_F/\Delta$
in the presence of interactions, there is a competition
between thermal suppression of coherent two-particle propagation in the wire
and activation of tunneling at the junctions at low temperatures.
As a result, the critical current shows
maximum as a function of temperature.
At even higher temperatures,
$k_B T  \gg \hbar v_{\rho}/2\pi d$, the suppression becomes
exponential.

In our model it is assumed that the superconducting electrodes
do not influence the  uniformity
of the potential along the quantum wire, since
they are separated from the wire by thick barriers.
It was argued in Ref.~\cite{Gogolin} that a non-uniform potential
in the wire will lead to an effective change of the boundary
conditions for the electronic wave function, which in turn could strongly
affect
our results. However, a recent calculation \cite{Maslov}
of the Josephson current through an interacting quantum wire
of {\it finite} length is in agreement with our results.
This indicates that the results obtained are robust
with respect to the specific choice of boundary conditions.
They rather describe generic properties of the superconductor-Luttinger liquid
system.

If a finite voltage $V$  is applied between the junctions, the AC Josephson
effect occurs.  The AC current acquires phase shift  proportional to the
distance between the tunnel junctions. Moreover, the amplitude of AC current
depends on the voltage in an oscillatory fashion due to spin-charge separation.
The corresponding period depends on the ratio of the velocities of the spin and
charge excitations in the LL.

A quantum wire closed to a loop (or quantum ring)
shows interesting parity effects.
Boundary conditions on the electronic wave functions result in a
discrete set of quantum numbers, related to the number of particles and
angular momentum. We showed how these numbers can be tuned by applying a
gate voltage and a magnetic flux, and calculated the corresponding dependence
of the critical current on these parameters.
This dependence shows a rich behavior, which can be detected in an
interference experiment employing a SQUID. We showed that the dependence is
robust to thermal fluctuations up to experimentally measurable temperatures.

\acknowledgments
We would like to thank I.L. Aleiner, L.I. Glazman, Yu.V. Nazarov, G. Sch\"on,
C. Winkelholz, and A.D. Zaikin for useful discussions.
The financial support of the European Community
(HCM-network CHRX-CT93-0136 and
HCM Fellowship ERB-CHBI-CT94-1474),
the Deutsche Forschungsgemeinschaft
through SFB 195, and the Netherlands Organization
for Scientific Research (NWO) is gratefully acknowledged.
The authors also acknowledge the kind hospitality of ISI-Torino
(Italy) where part of this work was done.

\begin{figure}
\caption{The geometries discussed in the text: (a) one-dimensional
wire connected to two superconductors by tunnel junctions.
The distance between the junctions is $d$. (b)
Ring with circumference $L$ connected to two superconductors by tunnel
junctions. The distance between the junctions is $L/2$, the ring is
threaded by a magnetic flux $\Phi$.}
\label{system}
\end{figure}

\begin{figure}
\caption{Relevant diagram for Josephson tunneling in the limiting
cases (a) $ v_F/d \ll \Delta$ and (b) $ v_F/d \gg \Delta$. The shaded area
indicates the electron-electron interaction.}
\label{diagram1}
\end{figure}

\begin{figure}
\caption{The critical current of the wire as a function
of the temperature ($t= Td/ 2 \pi v_F$) for various values of the interaction
strength $g_{\rho}= 1.0, 1.25, 1.75, 2.0$.}
\label{iwiretemp}
\end{figure}

\begin{figure}
\caption{The critical current through the ring (normalized to
$\displaystyle{\frac{2 \pi e v_F}{L}\frac{G_1 G_2}{(4e^2/\hbar)^2}} $) at a
fixed
value of the  flux
$f_{\Phi} = 0.2$ as a function of the gate voltage and the temperature ($t= TL/
\pi v_F$) in the non-interacting case.}
\label{iringmug2}
\end{figure}

\begin{figure}
\caption{The same as in the previous figure for the interacting case;
for this plot we
choose $g_{\rho}=1.75$ but the result is rather
generic for repulsive interaction (the critical current is
normalized to $\displaystyle{\frac{2 \pi e v_F}{L}\frac{G_1
G_2}{(4e^2/\hbar)^2}
}$).}
\label{iringmug175}
\end{figure}

\begin{figure}
\caption{The critical
current through the ring at a fixed value of the
gate voltage $f_{\mu} = 0.2$ as a function of
the flux and the temperature ($t= TL/  \pi v_F$) in the interacting case
$g_{\rho}=1.75$ (the critical current is
normalized to $\displaystyle{\frac{2 \pi e v_F}{L}\frac{G_1
G_2}{(4e^2/\hbar)^2}}
$).}
\label{iringfluxg175}
\end{figure}

\begin{figure}
\caption{Relevant diagrams for AC Josephson current: (a)
 $ v_F/d \gg \Delta$ and (b) $ v_F/d \ll \Delta$. The shaded area
indicates the electron-electron interaction.
\label{keldiagr1}}
\end{figure}

\begin{figure}
\caption{The voltage dependence of sinusoidal (solid line)
and cosinusoidal (dashed line) components and the phase (dotted line)
of AC Josephson current; (a) $g_{\rho} = 1.75$ and (b) $g_{\rho} = 1$.
\label{J_cs(V)}}
\end{figure}

\begin{figure}
\caption{The voltage dependence of the amplitude
of AC Josephson current. Here,
$g_{\rho} = 2, 1.75, 1, 0.75, 0.5, 0.25$ for the curves from
top to bottom at zero voltage.
\label{J_a(V)}}
\end{figure}

\begin{references}
\bibitem[\mbox{$\ast$}]{Permanent address.} Permanent address.
\bibitem{Nitta}J. Nitta, T. Azaki, H. Takayanagi, and K. Arai, Phys.
Rev. B {\bf 46}, 14286 (1992); A. Dimoulas, J.P. Heida, B.J. van Wees,
T.M. Klapwijk, W. v.d. Graaf, and G. Borghs, Phys. Rev. Lett. {\bf 74}, 602
(1995).
\bibitem{Kastalsky}A. Kastalsky, A.W. Kleinsasser, L.H. Greene, R. Bhat,
F.P. Milliken, and J.P. Harbison, Phys. Rev. Lett. {\bf 67}, 3026 (1991).
\bibitem{Andreev}A.F. Andreev, Zh. Eksp.Teor. Fiz. {\bf 46}, 1823 (1964)
[Sov. Phys. JETP {\bf 19}, 1228 (1964)].
\bibitem{Beenakkerrev}For a review see: C.W.J. Beenakker in {\em Mesoscopic
Quantum Physics}, edited by E. Akkermans,  G. Montambaux, and J.-L Pichard
(North Holland, Amsterdam) and references therein.
\bibitem{NATO-ARW}Several papers in {\em Mesoscopic Superconductivity},
edited by F.W.J. Hekking, G. Sch\"on, and D.V. Averin, Physica B {\bf 203}
No. 3 \& 4 (1994) deal with superconductor-semiconductor devices.
\bibitem{Beenakkerquant}C.W.J. Beenakker and H. van Houten, Phys. Rev.
Lett. {\bf 66}, 3056 (1991); A. Furusaki, H. Takayanagi, and M.
Tsukada, Phys. Rev. Lett. {\bf 67}, 132 (1991).
\bibitem{vanWees}B.J. van Wees,
P. de Vries, P. Magne\'e, and T.M. Klapwijk, Phys. Rev. Lett. {\bf 69}, 510
(1992).
\bibitem{Hekking} F.W.J. Hekking and Yu.V. Nazarov,
Phys. Rev. Lett. {\bf 71}, 1625 (1993); Phys. Rev. B {\bf 49}, 6847 (1994).
\bibitem{Aslamasov}L.G. Aslamazov, A.I. Larkin, and Yu.N. Ovchinnikov,
Zh. Eksp.Teor. Fiz. {\bf 55}, 323 (1968)
[Sov. Phys. JETP {\bf 28}, 171 (1969)].
\bibitem{Grabert}For a review see:
{\em Single Charge Tunneling},
edited by H. Grabert and M.H. Devoret
(Plenum Press, New York, 1992).
\bibitem{Bauernschmitt}R. Bauernschmitt, J. Siewert, Yu.V. Nazarov,
and A.A. Odintsov, Phys. Rev. B {\bf 49}, 4076 (1994).
\bibitem{Khmelnitskii}B.L. Altshuler, D.E. Khmelnitskii, and B.Z. Spivak,
Solid State Comm. {\bf 48}, 841 (1983).
\bibitem{Emery}For a review, see: V.J. Emery,
in {\em Highly Conducting One-Dimensional Solids}, edited by
J.T. Devreese, R.P. Evrard, and V.E. van Doren (Plenum, New York, 1979);
J. S\'olyom, Adv. Phys. {\bf 28}, 201 (1979); H.J. Schulz (unpublished).
\bibitem{Kane}C.L. Kane and M.P.A. Fisher, Phys. Rev. Lett. {\bf 68},
1220 (1992); Phys. Rev. B {\bf 46}, 15233 (1992).
\bibitem{MatveevGlazman}K.A. Matveev and L.I. Glazman, Phys. Rev. Lett. {\bf
70}, 990 (1993); K.A. Matveev, D. Yue, and L.I. Glazman, Phys. Rev. Lett. {\bf
71}, 3351 (1993).
\bibitem{Loss}D. Loss, Phys. Rev. Lett. {\bf 69},
343 (1992).
\bibitem{footnote1}These parity effects, ocurring in a normal system,
are quite different from the ones discussed in Ref.~\onlinecite{Bauernschmitt}
which are related to superconductivity.
\bibitem{Haldane}F.D.M. Haldane, Phys. Rev. Lett. {\bf 47},
1840 (1981); J. Phys. C {\bf 14}, 2585 (1981).
\bibitem{Byers} N. Byers and C.N. Yang, Phys. Rev. Lett. {\bf 7},
46 (1961).
\bibitem{Loss2}D. Loss and P. Goldbart,Phys. Rev. {\bf B43},
13762 (1991); S. Fujimoto and N. Kawakami, Phys. Rev. B {\bf 48}, 17406
(1993).
\bibitem{Goni}A.R. Go\~ni, A. Pinczuk, J.S. Weiner, J.M. Calleja, B.S. Dennis,
L.N. Pfeiffer, and K.W. West, Phys. Rev. Lett. {\bf 67}, 3298 (1991).
\bibitem{Tarucha} S. Tarucha, T. Honda, and T. Saku,
Solid State Commun. {\bf 94}, 413 (1995).
\bibitem{Fazio}A.A. Odintsov, R. Fazio, and F.W.J. Hekking in
Ref.~\onlinecite{NATO-ARW}; R. Fazio, F.W.J. Hekking, and A.A. Odintsov,
Phys. Rev. Lett. {\bf 74}, 1843 (1995).
\bibitem{Maslov}D.L. Maslov, M. Stone, P.M. Goldbart, and D. Loss,
(unpublished),  cond-mat/9508020.
\bibitem{Fisher} M.P.A. Fisher, Phys. Rev. B {\bf 49}, 14550 (1994).
\bibitem{footnote3} The
tunnel junctions are assumed to have
linear dimensions that are small compared to their separation $d$, but
large compared to the Fermi-wavelength $\lambda _F$.
\bibitem{Abrikosov}A.A Abrikosov, L.P. Gorkov, and I.E. Dzyaloshinski,
{\em Methods of Quantum Field Theory in Statistical Physics},
Dover, New York (1975).
\bibitem{footnote5}For BCS superconductors in the clean limit, the gap $\Delta$
is given by $v_{F,S} /\xi$, where $\xi $ is the correlation length. Therefore,
comparing energy scales means simply comparing $d$ and $\xi$, up to the ratio
$v_F /v_{F,S}$ of Fermi velocities in the 1D system and the superconductor.
\bibitem{Mailly}D. Mailly, C. Chapelier, and A. Benoit,
Phys. Rev. Lett. {\bf 70}, 2020 (1993).
\bibitem{Tinkham}M. Tinkham {\em Introduction to Superconductivity},
Krieger, Malabar, Florida (1980).
\bibitem{Keldish} L.V. Keldysh, Zh. Eksp. Teor Fiz. {\bf 47}, 1515 (1964)
[Sov. Phys.-JETP {\bf 20}, 1018 (1965)].
\bibitem{Kulik} I.O. Kulik and I.K. Yanson, {\em The Josephson
Effect in Superconducting Tunneling Structures}, Israel Program for
Scientific Translations, Jerusalem (1972).
\bibitem{Winkelholz}C. Winkelholz, R. Fazio, F.W.J. Hekking, and G. Sch\"on
(unpublished).
\bibitem{Faziorev}R. Fazio, F.W.J. Hekking, and A.A. Odintsov (unpublished).
\bibitem{Gogolin} M.Fabrizio and A. Gogolin, Phys. Rev. B (to be published).


\end{references}
\end{document}